# Understanding the mechanism of oxygen evolution reaction (OER) with the consideration of spin

Xiaoning Li, Zhenxiang Cheng* and Xiaolin Wang

*Institute for Superconducting & Electronic Materials (ISEM), Australia Institute for Innovative Materials, Innovation Campus, University of Wollongong, Squires Way, North Wollongong, NSW 2500, Australia.*

**Corresponding Author**

*E-mail: cheng@uow.edu.au

**Abstract**

Oxygen evolution reaction (OER) with intractable high overpotential is the rate-limiting step for rechargeable metal-air battery, water electrolysis systems, and solar fuels devices. There exists a spin state transition from spin singlet $OH^-/H_2O$ reactant to spin triplet $O_2$ product, which has not received enough attention yet. In this perspective, we attempt to retrospect electron behaviours during the whole OER process, with the consideration of spin attribute. Regardless of the adopted mechanisms by different electrocatalysts, for example, adsorbate evolution mechanism (AEM) or lattice oxygen mechanism (LOM), the underlying rationale is that active sites have to extract three in four electrons with the same spin direction before the formation of O=O. This spin-sensitive nature of OER superimposes additional high requirements on the electrocatalysts, especially on the spin structure, to compliment the fast electron transfer in the interface with spin selection and smoothly delivery afterwards. When optimizing the geometric and electronic structures catering for the spin-sensitive OER, awareness of the couplings between spin, charge, orbital and lattice is necessary. Some spin-correlated physical properties, such as (1) crystal field, (2) coordination, (3) oxidation, (4) bonding, (5) $e_g$ electron number, (6) conductivity and (7) magnetism, are also discussed briefly. It is hoped that our perspective could shed lights on the underlying physics of the slow kinetics of OER, providing a rational guidance for more effective energy conversion electrocatalysts designs.

## 1. Introduction

Despite nearly a century of research towards the oxygen evolution reaction (OER), its intractable sluggish kinetics still inhibit the developments of clean energy applications, such as rechargeable metal-air battery, water electrolysis systems, and solar fuels devices[1,2]. Different from the hydrogen evolution reaction (HER) [3], the sluggish kinetics of OER originates in the multiple intermediate steps to realize a four-electrons charge transfer as well as a singlet-to-triplet spin transition. An obstacle that badly impedes the fully understanding of OER mechanism is that the spin attribute of electrons involved has not been well recognized by the community [4]. The ground spin state of reactant $OH^-/H_2O$ is singlet without unpaired electrons, but contrastingly the product $O_2$ is a spin triplet with two unpaired electrons (Figure 1). In theory, the rate of a chemical reaction will be extremely slow if the spin of the electronic wave function of the products differs from those of the reactants, as the Hamiltonian does not contain spin operators.[5] Because it is forbidden by quantum mechanics, the oxygen evolution reaction is necessarily associated with an additional energy stimulus to proceed, such as from a magnetic field, thermal disturbance, or electrical potential [6]. Therefore, the goal of an OER electrocatalysts is to reduce the applied electrical potential for the $O_2$ molecular evolution, which claims at least two electrons to be extracted with the same spin direction. Therefore, to achieve a true breakthrough in diminishing the high overpotential of OER, more consideration needs to be given to the "spin transition" on top of "charge transfer".

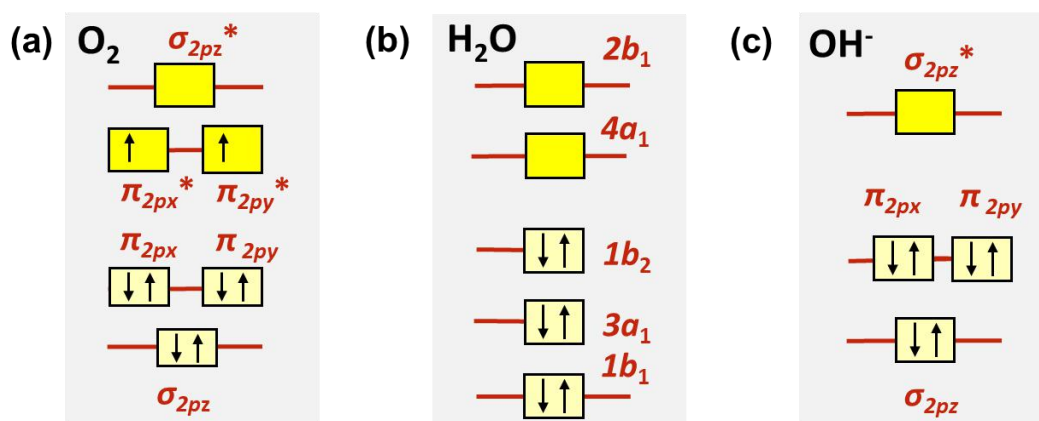

**Figure 1** Simplified molecular orbital of (a) $O_2$ with two unpaired electrons occupied the antibonding π* orbitals; (b) $H_2O$ with two electrons contributed by 2H; (c) $OH^-$ with two electrons contributed by 2H but losing one H orbital.

From the practical application point of view, most electrocatalysts are powders applied with conductive bindings, or thin film coated or grown on electrodes during the OER tests[7-9]. Electrons extracted from absorbates by electrocatalyst surface have to pass through the electrocatalysts body before flowing into the external circuit, even for epitaxial surfaces with 15 nm thickness [10]. It is well recognized that the oxygen reduction efficiency is related with bulk conductivity and charge transfer[11]. Therefore, the overpotential applied actually arises from two parts with different physical mechanisms. The primary part is the energy barrier for surface reaction (SR) associated with electron transfer from absorbates to the active sites. The other ineluctable part is the energy consumed by electron bulk transport (BT) in electrocatalyst body. With the consideration of spin, ideal electrocatalysts should endow the capability to extract at least two electrons in the same spin direction before the formation of O=O (SR), and simultaneously ensure a rapid transport of these spin polarized electrons (BT)[4,12]. To date, most OER electrocatalysts are still constrained by the primary surface reactions part (SR) for the inefficiency for fast electron transfer with spin conversation, which kinetics is so slow that the subsequent bulk transport part (BT) seems relatively trivial. Even though, many cations such as $Fe^{3+}$ are high OER-active on the surface but its performance are rigorously limited by poor electrical conductivity of corresponding compounds, and a good bulk conductivity could lead to a larger fraction of active sites participating in catalysis [13]. It suggests that promoting the bulk transport (BT) of spin-polarized electrons that extracted from the surface is quite important to guarantee a fast kinetic of surface reaction (SR) to extract these electrons. Hence, unveiling the electron behaviours during the entire process of OER with full consideration of spin is the key to open the door of the true OER mechanism.

In this perspective, we will firstly retrospect the surface reaction (SR) with the consideration of spin based on various OER mechanisms proposed previously, to disclose the pivotal role of spin plays. Then bulk transport (BT) is also ruminated for the case of electrons with polarized spin to abate the superfluous energy consumption. Rationales for optimizing electrocatalysts to meet the requirement of spin is addressed based on $3d$ transition metal oxides.

## 2. Retrospect OER with the consideration of spin

The transformation from $OH^-$ to $O_2$ in alkaline or $H_2O$ to $O_2$ in acid medium has similar intermediate oxygen species and evolution process for various widely accepted reaction mechanisms [14]. OER can be treated as a process of depletion of valence electrons of oxygen in the reactant $OH^-/ H_2O$ (0) with 8 valence electrons per oxygen atom that satisfies the octet rule, to release product $O_2$ (5) with 6 valence electrons, among which two occupied π* are unpaired with the same spin direction. From $OH^-/ H_2O$ (0) to $O_2$ (5), the intermediates *OH (1), *O (2), *OOH (3), *OO (4) are chemisorptively bonded with active sites in the electrolyte-electrocatalyst interface, which are intrinsically electron deficient due to the charge transfer to the active sites [15]. Figure 2 provides three basic reaction paths of OER process concluded among various electrocatalysts, based on which the role of spin will be elucidated thoroughly.

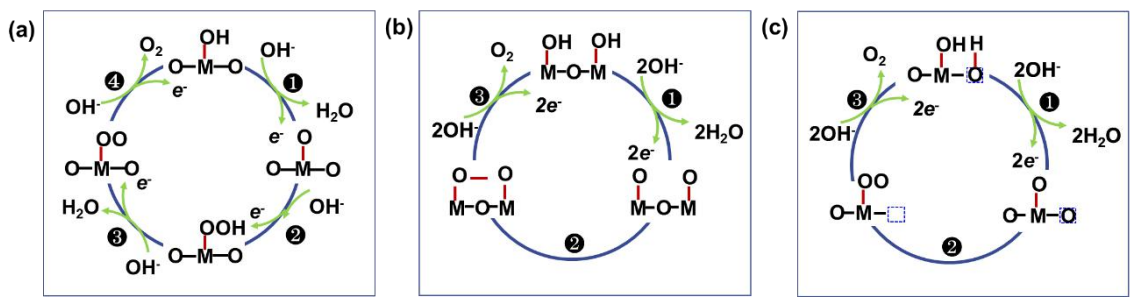

**Figure 2** Well-recognized reaction paths proposed based on various electrocatalyst working in alkaline OER electrolytes. (a) Eley–Rideal (ER) type adsorbate evolution mechanism (AEM) with single metal cation active site, from reactant OH⁻(0) to intermediates *OH (1), *O (2), *OOH (3), *OO (4), to product $O_2$ (5). The number after each oxygen species is marked for a better logics in main text, but not marked in figures to avoid intricacy; (b) Langmuir–Hinshelwood (LH) type adsorbate evolution mechanism (AEM) with two adjacent metal cation active sites; (c) lattice oxygen mechanism (LOM) with one metal cation active site and one oxygen in lattice involved. LH-tpye AEM and LOM is with reactant OH⁻(0) to intermediates *OH (1), *O (2), *OO* (4), to product $O_2$ (5), with the bypassing of *OOH (3).

In most cases, different reaction routes will be favoured by different electrocatalysts (or by different local crystal plane surfaces in the same electrocatalyst) in order to minimize free energy of activation. For instance, there are two different reaction paths based on adsorbate evolution mechanism (AEM): (1) the Eley–Rideal (ER) type and (2) the Langmuir–Hinshelwood (LH) type [16]. (Figure 2a, b). Adsorption of reactants or intermediates on the surface of electrocatalysts is actually the results of orbital overlap, generation of chemisorptive bond, and charge transfer in the interface. The real depletion of an electron from adsorbate to the active site proceeds only when it has enough energy under applied potential. Figure 3a shows one possible situation of the outer electron's behaviours during the ER-type AEM process. The six electrons of OH⁻ (0) in electrolyte are all paired occupied the three low energy orbitals, from which the active site will select spin-down one for the formation of *OH (1). In this stage, the electron actually is still shared by the adsorbate and surface-active site and it should be extracted from surface via bulk transport to the external circuit to make the active site available for the transformation of *OH (1)-*O (2). If the electron extracted during the formation of *OH (1)-*O (2) is spin down, and the next one during the formation of *O (2)-*OOH (3) is spin up, and the subsequent one during the formation of *OOH (3)-*OO (4) should be spin down for the evolution of $O_2$ (5) with lowest energy principle. If the electron extracted during the formation of *OH (1)-*O (2) is spin down, but the next one during the formation of *O (2)-*OOH (3) is also spin down, then the subsequent two electrons should be with different spin directions. Many cases can be possible since the spin selection is dictated by contemporary spin configuration and interaction of adjacent active sites, but the rationale is that three in four electrons are in the same spin direction.

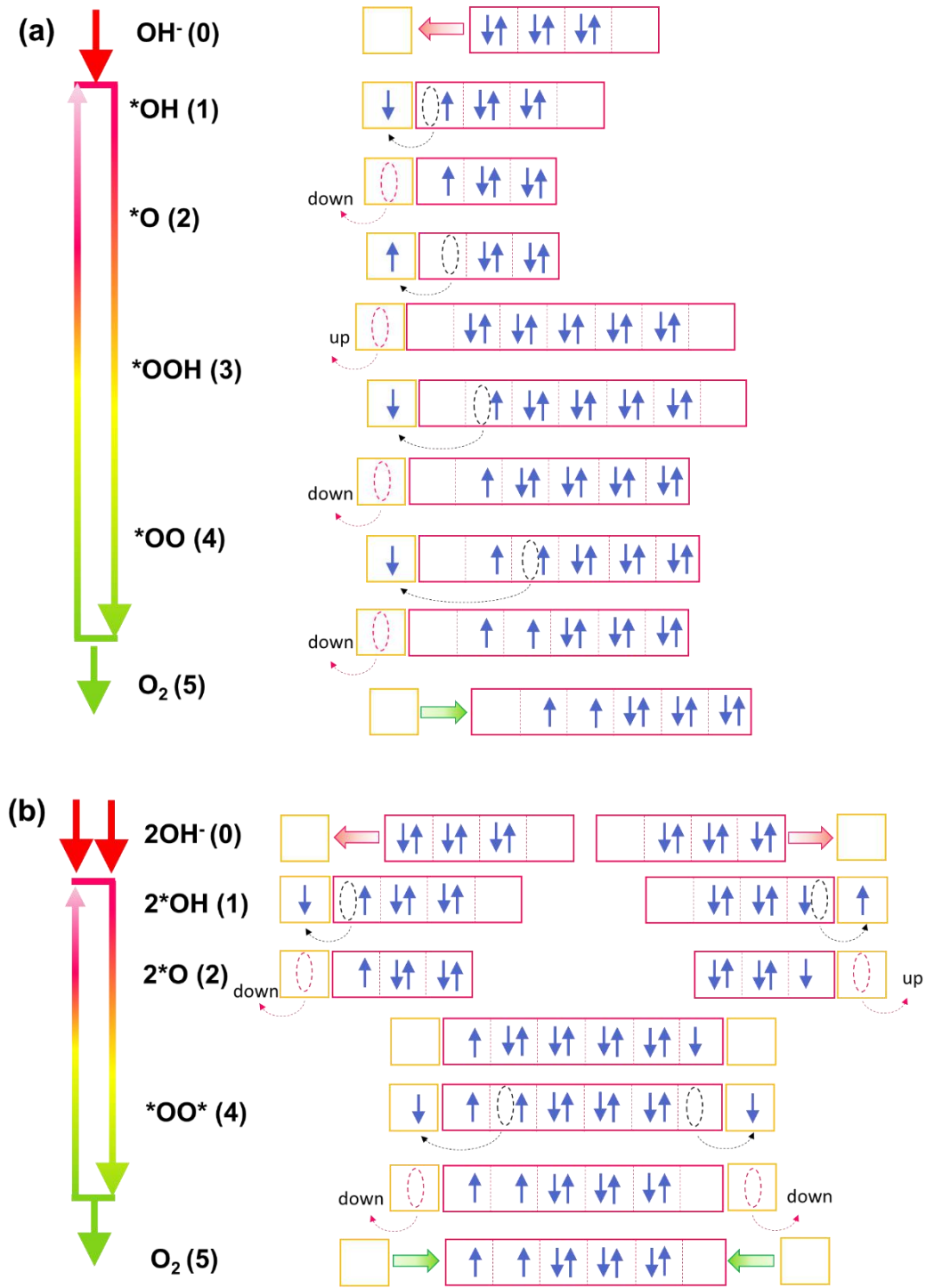

**Figure 3** Possible cases of the outer electron's behaviours during OER process. (a) is the cases for one active site mechanism such as ER-type AEM; (b) is the cases for two active sites mechanism such as LH-type AEM and LOM. The red squares present the hybrid orbitals of oxygen species, which is simplified without the energy level contrasting to Figure 1. The black dashed circles and lines with arrows track the electron transfer/hopping between adsorbates and active sites via chemisorptive bonding. The electron depletion occurs as the red dashed circles and lines denoted, which are delivered away surface to bulk. Here, the spin

direction of the four electrons are down-up-down-down, other possible cases can also be possible if three in four electrons are in the same spin direction.

For the situation of LH-type AEM, in which the transformation of *O (2)-*OOH (3)/ *OOH (3)-*OO (4) in ER-type reaction is replaced by *O (2) -*OO* (4), and the energy barrier of which is relatively lower as it detours the rate-limiting step of OOH* (3) with the collaboration of two adjacent cation active sites (Figure 2b). For instance, the OER activity of quadruple perovskite $CaCu_3Fe_4O_{12}$ (CCFO) exceeds that of the benchmark $Ba_{0.5}Sr_{0.5}Co_{0.8}Fe_{0.2}O_{3-\delta}$ (BSCF) and $RuO_2$ by adopting the LH-type mechanism [17]. As explained, it is benefited from the catering geometric structure of the shorten oxygen–oxygen distance (~2.6 Å) induced by heavily bent Fe–O–Fe bonds. Lately, O-O formation via direct coupling between metal cation and lattice anion known as lattice oxygen mechanism (LOM) is observed in today's most active transition metal oxide OER electrocatalysts [18-21]. The underlying physics of LOM is that the Fermi level is lowered into oxygen $2p$ band induced by the strong overlap of metal $3d$ with oxygen $2p$ [22]. Generally, the kinetic of LOM is even faster than ER-type AEM due to the non-concerted proton–electron transfer process, as reported in LOM-adopted $CaCoO_3$ with substantially smaller lattice parameter and shorter surface oxygen separation [23]. In this situation, the redox centre of the catalyst is no longer limited to the metal alone, and the ligand holes are also involved in the oxidation reaction (Figure 2c). The spin of ligand holes may be coupled with that of adjacent metal, providing a shortcut for the formation of spin triplet $O_2$. It's noteworthy that the prerequisite for rapid LOM reaction is that O $p$-band centre is close enough to the Fermi level, which is also the culprit for poor stability of electrocatalysts [24].

One of the differences between LH-type AEM and LOM is that lattice oxygen is activated as active sites in LOM. Similarly, the rationales based on these two active sites involved mechanisms is in line with that concluded from one active site mechanism, as depicted in Figure 3b.

## 3. General principle of OER with the consideration of spin

With a clear understanding of the spin-sensitive OER process, that is, 3 in 4 electrons are in the same spin direction, some well-acknowledged principles proposed previously would be much reasonable, concerning surface adsorption and bulk conductivity.

Sabatier principle suggests that the adsorptive interactions between the catalyst and the adsorbates are expected to be "just right" [25]. Adjusting adsorption capacity of active sites has been regarded as one of the most powerful strategies to reduce the surface reaction energy barrier. For 3$d$ transition metal based oxides with octahedron coordination, the adsorption capacity correlates to the $e_g$ occupation, due to spatial overlap and energetic similarity of the electronic states [16,26]. Yang Shao-Horn's research group revealed that the OER activity exhibited a volcano-shaped dependence on the number of $e_g$ electrons of transition metal perovskite oxides [27]. The relatively fast kinetics of $Ba_{0.5}Sr_{0.5}Co_{0.8}Fe_{0.2}O_{3-\delta}$ (BSCF) with $(t_{2g})^5(e_g)^{1.2}$ is boosted by the shift of rate-limiting step from OH$^-$(0)-*OH (1) to *O (2)-OOH* (3) [28]. In fact, near unitary $e_g$ occupation provides empty one $e_g$ orbital, which make it easier for active site to select one electron from OH$^-$ (0) with spin paralleled with that of $e_g$ electron for the formation of *OH (1) without pairing. Moreover, it's also easier to be transported to the adjacent bulk cations without de-pairing. Nevertheless, the complexity of the local surface, such as reconstructions, space charges, polarity or segregation [29], also hampers the observation of OER mechanism for particular electrocatalysts, meriting more experimental and computational study with the spin consideration.

Meanwhile, rapid delivery away and transport to external circuit of these extracted electrons cannot be disregarded to warrant the availability of surface-active sites. Intuitively, good electrical conductivity of electrocatalysts can effectively reduce the interfacial and bulk transport resistances, resulting in enhanced OER performances as many researchers have already suggested [30,31]. However, there are also other reports indicating that enhancing bulk conductivity is neither sufficient nor necessary for obtaining a better OER performance[32,33]. This is appreciable in view of the fact that 3/4 incoming electrons are in the same spin direction. To rapidly transport these spin polarized electrons with theoretical minimum resistance, electrocatalysts with half-metallic peculiarity with high spin polarisation at the Fermi level ($E_F$) may be a better choice. In general, the conductivity of 3$d$ transition metal (hydro)oxides is not so good as noble-metal based compounds, but they possess rich physical properties with large regulation freedom due to the couplings of spin-charge-orbital-lattice. They can be half-metallic, metallic, semiconducting, or Mott insulating if with deliberate manipulation [34]. That's one of the reasons for the superiority of some 3$d$ transition metal based electrocatalysts with half-metallic conductivity [35].

In short, spin state transition from singlet reactant to triplet product is responsible for the high overpotential of OER, in any proposed mechanisms, either AEM or LOM. Accelerating the transfer and transport of spin-polarized electrons is the core task for OER electrocatalysts.

Except for a mediated adsorption/desorption strength for rapid charge transfer, spin-sensitive OER imposes additional requirements on the spin selection and spin conductivity.

## 4. Spin-correlated factors that influences OER in electrocatalysts

Due to the sensitivity of spin during the OER process, electrocatalyst that can efficiently accelerate this process should possess very catering spin configurations with deliberately modulation based on a fully understanding of the OER process with the perspective discussed above. When optimizing the geometric and electronic structures in the real electrocatalysts, awareness of the couplings between spin, charge, orbital and lattice is strongly necessary. Generally, spin configuration of an electrocatalyst is mainly decided by crystal structure and element composition. The original crystal structure, hosting for active sits, determines the basic framework of potential coordination environment, general bond length, and thus the average crystal field splitting. Then, a group of different elements occupying the given crystal lattice would lead to specific oxidation states and thus the spin configuration [36]. It should be pointed out that the elements composition may also affect the crystal structure reversely, due to the couplings of spin, charge, orbital, and lattice parameters. The extremely complexity of the coupling of these factors are summarized in Figure 4, which we have tried best to simply with some missing details.  It's not quite accurate to conclude that OER performance is positively or negatively correlated with one specific factor, which may vary case by case. However, the ultimate effect of these factors on the catalyst is the ad-/de- sorption process and the polarized electron transport. Herein, we only briefly discuss several spin-correlated physical properties to provide some hints for effective regulation.

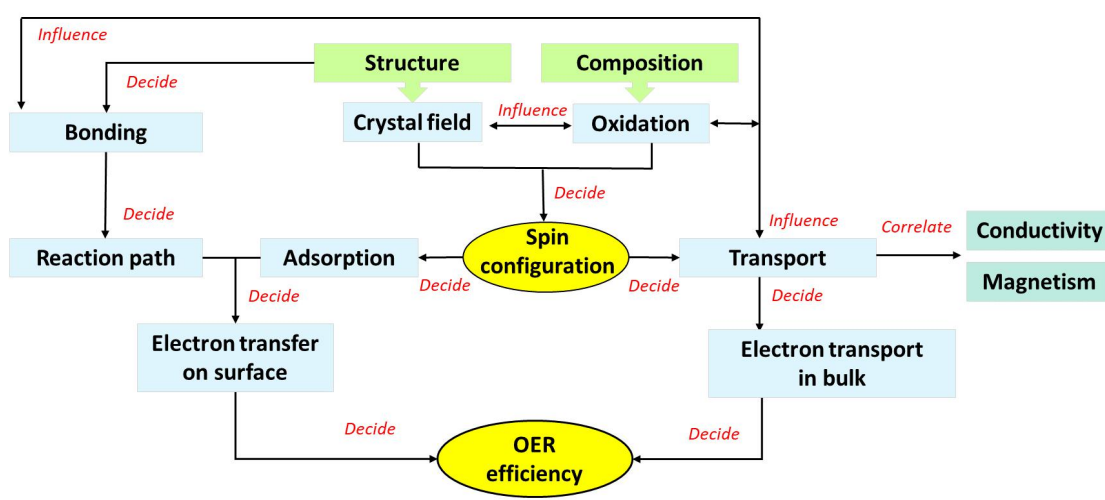

**Figure 4** Demonstration of the relationships of spin-correlated factors influenced OER. Basically, OER efficiency of an electrocatalysts is decided by electron transfer on the surface and electron transport in bulk, both of which is based on the spin configuration of the electrocatalyst. The spin configuration can be adjusted from the crystal structure and composition, which is directly determine the crystal field and oxidation state. However, crystal structure and composition variation also impose effects on each other due to couplings of fundamental physics. The structure of an electrocatalyst with different bonding length or angle can also change the OER intermediates reaction path. In terms of electron transport in bulk during OER process, it also correlates with other physical properties of the materials, such as conductivity and magnetism. Therefore, intelligent manipulation of electrocatalysts for substantial improve OER performance is still a big challenge.

**4.1 Crystal field splitting.** Crystal field splitting plays a very important role in determining the spin configuration of $d$-band transition metals. In principle, electrons tend to have the maximum unpaired number to reduce the pairing interaction energy ($P$), but additional energy is required to occupy the high energy orbital in order to overcome the crystal field splitting energy ($\Delta$). In 3$d$ transition metal oxides, the $P$ and $\Delta$ are sometimes of comparable magnitude, which can lead to competing spin states [37]. Crystal field splitting energy ($\Delta$) is associated with coordination, bonding, oxidation state, $e_g$ occupation and ligands [38-40].

**4.2 Coordination.** Different coordination will lead to disparate spin configurations. For instance, five-fold 3$d$ orbitals in an octahedral symmetry will be split into two sets with different energy: three lower energy $t_{2g}$ orbitals and two higher energy $e_g$ orbitals. But for a tetrahedral crystal field, the situation will be two lower energy $e_g$ and three higher energy $t_{2g}$ states. The influence of transition metal coordination on the OER activity and stability of perovskites has been reported by many researchers [38].

**4.3 Bonding.** Bond length, bond angle, and atomic distance not only can control the reaction path adopted by surface reaction as mentioned above, but also can influence the interactions of adjacent ions. Increasing the covalent metal-oxygen bonding increases the magnetization of the oxygen atoms through spin donation from certain high-spin transition metal ions. The monotonic increase in OER activity associated with an average Mn−Mn distance from ~3 to ~3.2 Å was observed in the manganese oxides $Mn_2O_3$ and $Mn_3O_4$, and in $AMnO_3$ and $AMn_7O_{12}$ perovskites (A = Ca, La, etc.) [41]. Metal-oxygen-metal bonds bending could decrease the oxygen 2p−metal 3d transfer energy value, resulting in a higher OER efficiency [42].

**4.4 Oxidation states.** Higher oxidation state means less electrons are left in the $d$ orbitals after combining with an oxygen ligand. For example, a $Co^{2+}$ ($3d^7$) octahedron has two possibilities: high spin (HS) $(t_{2g})^5(e_g)^2$ and low spin (LS) $(t_{2g})^6(e_g)^1$. While for $Co^{3+}$ ($3d^6$) with one electron less, the spin state can be HS $(t_{2g})^4(e_g)^2$, LS $(t_{2g})^6(e_g)^0$, or intermediate spin (IS) $(t_{2g})^5(e_g)^1$. Sometimes a non-integer electron number measured for a certain compound can also be possible due to co-existing of cations with different oxidation state or a high degree of hybridization (charge transfer) between transition metal cations and oxygen anions [43,44]. It should be noted that high oxidation state would result in a smaller ion radius, shorten bond length, generation of oxygen vacancies, ligand hole and so on [21,43,45].

**4.5 $e_g$ occupation.** In the octahedron, $e_g$ orbitals will have strong spatial overlap with those of adjacent O 2p orbitals, forming σ-bonding and σ*-antibonding states with partial metal and oxygen character. The presence of a single/unit electron in the antibonding $e_g$ orbital is expected to yield just the appropriate strength of interaction. While uneven occupation of near unitary $e_g$ electron induces Jahn–Teller distortion which would elongate the lattice and weaken the metal-oxygen covalent bonds along $c$-direction [46].

**4.6 Conductivity.** Transport property of electrocatalyst can partially reflects the capacity to conduct electrons with certain spin. Electrons of electrocatalyst with strong orbital overlap of metal 3d-oxygen 2p are partially delocalized losing its orbital angular momentum, thus with high spin conductivity (metallic like). If the overlap is weak, they communicate electrons through exchange interactions, such as super-exchange interaction and double-exchange interaction, as explained by the Goodenough-Kanamori rules[47]. Double-exchange interaction usually involves electron hopping with conserved spins, resulting a ferromagnetic alignment of the electrons from two adjacent metal cations (half-metallic like) [48].

**4.7 Magnetism.** Also rooting in the exchange interaction, magnetism of electrocatalysts could be instructive to the possible electron transfer. For example, most transition metal oxides with LOM are paramagnetic (PM)materials without exchange interaction due to the strong orbital overlap partially losing orbital and spin angular momentum [49]. The well-recognized highly-efficient OER electrocatalyst $Ba_{0.5}Sr_{0.5}Co_{0.8}Fe_{0.2}O_{3-\delta}$ (BSCF) has an A-type antiferromagnetic (A-AFM) ground state with spin accumulation in the ferromagnetic (FM) plane.[4] In contrast, G-type antiferromagnetic (G-type AFM) ordered oxides are not-so-good OER electrocatalysts, since spin is cancelled out completely in any directions [50]. A-type ferrimagnetis (A-FiM) materials with uncompensated magnetic moments (FM ordering intra-layer and AFM inter-layer) are also a good OER material family [51,52].

## 5. Conclusions

The intractable high overpotential of OER is associated with the spin state transition from reactant to product, which can be curtailed if 3/4 electrons extracted from adsorbates by surface active sites are with the same spin direction and can be speedily transported afterwards. Although this conclusion is made in the absence of substantial input, but it is crucial in the understanding of intrinsic OER mechanism. Bearing this in mind, it is viable to discern the role of spin on the OER intrinsically based on deliberate designs of comparative experiments together with advanced characterization techniques. Specifically, application of synchrotron radiation, spherical aberration corrected electron microscope is powerful to probe the local geometric and electronic structure. To discern the capability of a material to conduct a spin current, resistance measurements under a magnetic field to align the spin may be a good method. Fortunately, electric transport or magnetic property can also be referred to when selecting potential electrocatalyst material systems, for the similar mechanisms rooting in the electron transfer affected by spin-charge-orbital-lattice couplings. The role of parallel spin has been experimentally justified by applying an external magnetic field to ferromagnetic electrocatalysts, which favours the parallel alignment of spin during the formation of the O–O bond, resulting in significant enhancement of OER efficiency (up to 40%) [53]. Therefore, it wouldn't be long to realize the substantially reduced OER overpotential, with adequate understanding of OER with full solicitude of spin and a deliberate regulation on the structure, to meet the requirement of the targeted energy conversion applications.

**Data availability**

Data available from the corresponding authors upon request.

**Competing financial interests**

The authors declare no competing interest.

**Acknowledgement**

This work was financially supported by Australia Research Council (DP190100150, DP170104116). We thank Dr. Tania Sliver for the critical reading about the manuscript.

**Author contributions**

Z.C., and X.L. conceived the theme. X.L. wrote the manuscript and designed the schemes and figures. Z.C. and X.W contributed with insights, discussions and modification on the manuscript.